\documentclass[aps,pre,twocolumn,superscriptaddress,showpacs]{revtex4-1}
\usepackage{epsfig}
\usepackage{amsmath,amssymb,bm}
\usepackage{graphics}
\usepackage{epsfig}
\usepackage{graphicx}
\begin{document}
\title{
Giant components in directed multiplex networks
}
\author{N. Azimi-Tafreshi}
\affiliation{Physics Department, Institute for Advanced Studies in Basic Sciences, 45195-1159 Zanjan, Iran }
\author{S.~N. Dorogovtsev}
\affiliation{Departamento de F{\'\i}sica da Universidade de Aveiro $\&$ I3N, Campus Universit\'ario de Santiago, 3810-193 Aveiro, Portugal}
\affiliation{A.F. Ioffe Physico-Technical Institute, 194021 St. Petersburg, Russia}
\author{J.~F.~F. Mendes}
\affiliation{Departamento de F{\'\i}sica da Universidade de Aveiro $\&$ I3N, Campus Universit\'ario de Santiago, 3810-193 Aveiro, Portugal}
\begin{abstract}
We describe the
complex global structure of giant components in directed multiplex networks which generalizes the well-known bow-tie
structure, generic  for ordinary directed networks.
By definition, a directed multiplex network contains vertices of one type and directed edges of $m$ different types.
In directed multiplex networks, we distinguish a set of different giant components based on the existence of directed paths of different types between their vertices, such that for each type of edges, the paths run entirely through only edges of that type.
If, in particular, $m=2$, we define a strongly viable component as a set of vertices, in which for each type of edges, each two vertices are interconnected by at least two directed paths in both directions, running through the edges of only this type. We show that in this case, a directed multiplex network contains, in total, $9$ different giant components including the strongly viable component. In general, the total number of giant components is  $3^m$.
For uncorrelated directed multiplex networks, we obtain exactly the size and the emergence point of the strongly viable component and estimate the sizes of other giant components.
\end{abstract}
\maketitle

\section{Introduction}
The so-called bow-tie organization of giant components is a generic feature of directed networks \cite{BKMRRSTW,NewmanStWatts,DoroMendessamukhin}. This structure was first reported for the directed WWW graph, but it is actually valid for general directed networks, see Fig.~\ref{f1}(a). There are three different giant (i.e., containing a finite fraction of all vertices in an infinite network) components in these networks. From each vertex of the giant strongly connected component, S, one can reach any other vertex by a directed path, so that each two vertices are mutually reachable. From each vertex of the giant in-component, I, one can reach the vertices of the strongly connected component by a directed path. Each vertex of the giant out-component, O, is reachable from vertices of the strongly connected component. (Note that according to this definition, the strongly connected component is included both in the giant in- and out-components.) In uncorrelated networks and, more generally, in locally tree-like networks with given degree-degree correlations, the sizes of these giant components were obtained analytically \cite{NewmanStWatts,DoroMendessamukhin,Baguna}.

In recent few years, the focus of interest of the complex networks studies essentially shifted from single networks to coupled networks, networks of networks, etc., including interdependent and multiplex networks \cite{GaoBuldStanleyHalvin,ParshaniBuldHalvin,BuldParshaniPaulStanley,Vespignani,
SonChrisGrassPaczuski,BaxterDoroGoltsev,HuKshCohenHavlin,mathematicalmultiplex,ML11}. In the interdependent networks, each vertex in a network depends on a vertex or several vertices in other networks. As a result, removal of vertices in one network may launch a cascade of failures destroying a finite fraction of all networks \cite{BuldParshaniPaulStanley}. Depending on the structure of these networks and the fraction of the initially removed vertices, this cascade may eliminate the networks completely or remain a finite fraction of nodes and edges undamaged \cite{Vespignani,SonChrisGrassPaczuski,BaxterDoroGoltsev}. The specific phase transition between these two situations is hybrid, which means that it combines a discontinuity and a critical singularity \cite{HuKshCohenHavlin}. In the simplest representative situation, in which each vertex in interdependent networks has not more than one interdependence, the interdependent networks are actually equivalent to multiplex networks \cite{Mucha,KABGleeson,mathematicalmultiplex,ML11,Bianconi}. The multiplex networks have vertices of one type and edges of several different types. Hence, multiplex networks are graphs with edges of several different colors. They can be treated as a superpositions of several graphs of distinct colors. The role of the remaining giant component of interdependent networks, in multiplex networks, plays the giant viable cluster. For each type of edges in an undirected multiplex network, each two vertices in the viable cluster are connected by at least one path following edges of that type (i.e., there must be at least one path of each color between each two vertices) \cite{SonChrisGrassPaczuski,BaxterDoroGoltsev}.

Previously, undirected multiplex networks were explored \cite{BaxterDoroGoltsev,mathematicalmultiplex,ML11, Mucha,KABGleeson,Bianconi,BaxtDoroMenCellai,Overlap,BianconiDorogovtsev,MinLeeGoh,AzimiGomezDoro,Halu:hmb14}. In the present article, we study multiplex networks, in which all edges are directed. We show that the giant components in these networks are organized in an essentially more complicated way than in ordinary directed networks. We introduce a system of different giant components based on the set of
directed paths of different types connecting the vertices in these components. Figure~\ref{f1}(b) demonstrates this set of giant components in directed multiplex networks with edges of two types. For locally tree-like networks, we find the size of strongly viable component analytically and describe its simplest structural characteristics and the hybrid phase transition associated with the emergence of this component. For directed multiplex networks with two types of edges, we obtain lower limit estimates for the sizes of other viable components.

This paper is organized as follows.
In Section~\ref{directed networks sec}, we introduce the set of giant viable components in directed multiplex networks, including the strongly viable component. We derive equations enabling us to describe these components, if the networks are locally tree-like, in Sec.~\ref{equations}. In Section~IV we analyse the case of directed multiplex networks with two types of edges and find the sizes and the hybrid phase transition associated with the emergence of these components. We  also show how to find the fraction of edges belonging to the strongly viable component.


\section{Giant components in directed multiplex networks}
\label{directed networks sec}
Let us
introduce different giant viable components in directed multiplex networks. For the sake of brevity, we define them in the particular case of multiplex networks having two types of edges, i.e., edges of two colors, $A$ and $B$. This network can be treated as a superposition of networks with edges of color $A$ (network $A$) and the network with edges of color $B$ (network $B$). Generalization to the case of an arbitrary number $m$ types of edges is straightforward. Our definition is based on interconnectivity of different parts of these networks, understood in terms of the set of different directed paths running between these parts. One can introduce directed paths of each color, i.e., the paths following directed edges of only that color.
\begin{figure}
\begin{center}
\scalebox{0.24}{\includegraphics[angle=0]{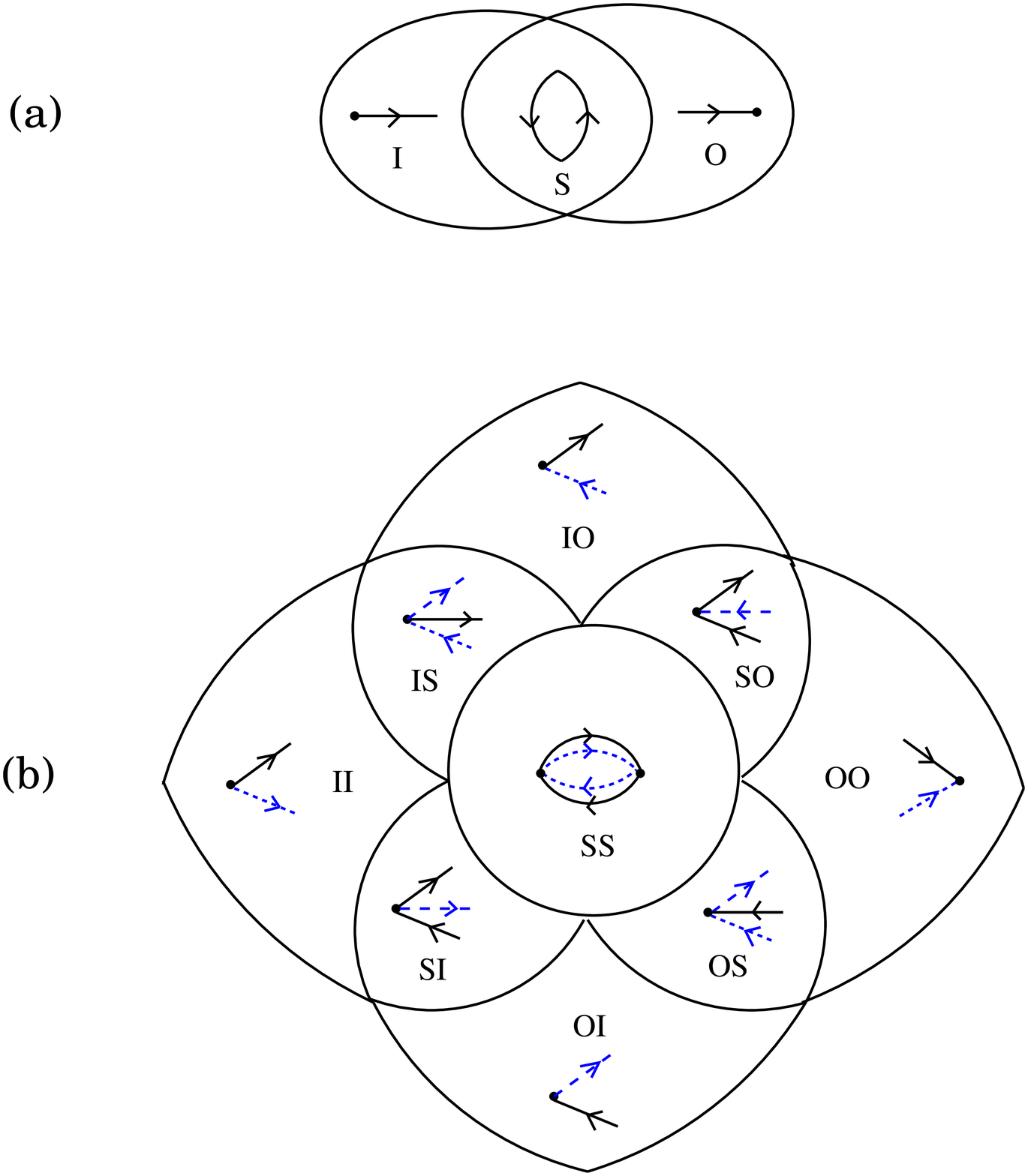}}
\end{center}
\caption{(Color online) The structure of giant components
in (a) ordinary directed networks and (b) in multiplex directed networks.
}
\label{f1}
\end{figure}
\begin{figure*}
\begin{center}
\scalebox{0.44}{\includegraphics[angle=0]{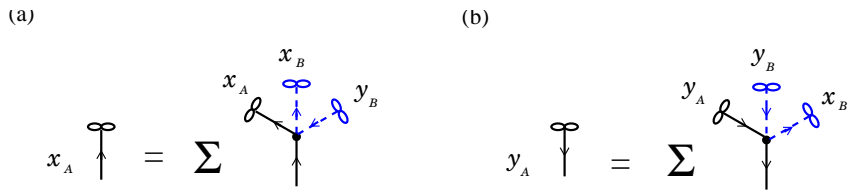}}
\end{center}
\caption{(Color online) Schematic representation of the self-consistency equations for the probabilities $(a)$ $x_A$ and
$(b)$ $y_A$, $m=2$. The solid black and dashed blue lines with infinity symbols at one of their ends, represent probabilities $x_A,y_A$ and $x_B,y_B$ respectively. The right-hand sides of these equations indicate the full sets of configurations contributing to the corresponding probabilities.
for the sake of clarity, we do not show the edges leading to finite components.
}
\label{f2}
\end{figure*}

(1)
We define the strongly viable connected component, SS, in the following way. Each two vertices in this component
are reachable from each other by directed paths of both colors.
Clearly, SS is a subgraph of the giant strongly connected components of two networks with edges of distinct colors.

(2) From any vertex of the in-in viable component, II, there are at least two directed paths (a path of one color and a  path of the second color) to the strongly viable component SS. The in-in viable component is a subgraph of the  in-components of the two one-color networks.

(3) To any vertex of the out-out viable component, OO, there are at least two directed paths of both colors from the SS component. The out-out viable component is a subgraph of the out-components of the two one-color networks.

(4) From each of the vertices of the strongly-in viable component, SI, there are at least two directed (at least one in each direction) paths of color $A$ to SS and at least one directed path of color $B$ to SS. The strongly-in viable component is a subgraph of the strongly connected component of network $A$ and the in-component of network $B$.

(5) From each of the vertices of the in-strongly viable component, IS, there are at least two directed (at least one in each direction) paths of color $B$ to SS and at least one directed path of color $A$ to SS. The in-strongly viable component is a subgraph of the strongly connected component of network $B$ and the in-component of network $A$.

(6) To each of the vertices of the strongly-out viable component, SO, there are at least two directed (at least one in each direction) paths of color $A$ to SS and at least one directed path of color $B$ from SS. The strongly-out viable component is a subgraph of the strongly connected component of network $A$ and the out-component of network $B$.

(7) To each of the vertices of the out-strongly viable component, OS, there are at least two directed (at least one in each direction) paths of color $B$ from SS and at least one directed path of color $A$ from SS. The out-strongly viable component is a subgraph of the strongly connected component of network $B$ and the out-component of network $A$.

(8) From each of the vertices of the in-out viable component, IO, there is at least one directed path of color $A$ to SS and from SS there is at least one directed path of color $B$ to each of vertices of IO. The in-out viable component is a subgraph of the in-component of the network $A$ and the out-component of the network $B$.

(9) From each of the vertices of the out-in viable component, OI, there is at least one directed path of color $B$ to SS and from SS there is at least one directed path of color $A$ to each of vertices of OI. The out-in viable component is a subgraph of the out-component of the network $A$ and the in-component of the network $B$.

Note that according to these definitions, SS is a subgraph of the remaining giant components.
These nine components are schematically shown in Fig.~\ref{f1}(b). The number of the components exponentially grows with the number of colors, $m$, see below.
\begin{figure}[t]
\begin{center}
\scalebox{0.42}{\includegraphics[angle=0]{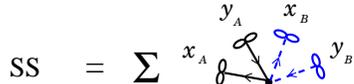}}
\end{center}
\caption{(Color online) Schematic representation of a vertex belonging to the giant strongly viable component.
}
\label{f3}
\end{figure}
\section{Equations for tree-like directed multiplex networks}
\label{equations}

The locally tree-like structure of random networks allows a simple analytical treatment
based on the generating function technique \cite{NewmanStWatts}. Let us consider locally tree-like directed multiplex networks with edges of $m$ types, labeled with $i=A,B,\ldots$. For the sake of simplicity, here we only consider the directed multiplex networks, in which each of
networks $i$ are uncorrelated although the numbers of different connections of a vertex may be correlated. This multiplex network is locally tree-like, that is, it has no finite loops (cycles in terms of graph theory) in the infinite size limit. Introducing the notation $\textbf{q}_{i}=(q_{in,i},q_{out,i})$ for the vector of the numbers of incoming and outgoing connections of a vertex in network $i$, we describe this multiplex network completely by the joint degree distribution $P(\textbf{q}_A,\textbf{q}_B,\ldots)$.

To find the relative size of the giant viable components, we introduce a set of probabilities which are defined in Fig.~\ref{f2} [the probabilities are shown only for one type of edges ($i=A$)].
$x_i$ is the probability that if we choose uniformly at random an edge of type $i$ and move along its direction, then the end vertex of this edge is the root of an infinite subtree each of whose vertices (including the root vertex) has at least one outgoing edge and one incoming edge of each type, see Fig.~\ref{f2}(a).
Similarly, $y_i$ is the probability that if we choose uniformly at random an edge of type $i$ and move against its direction, then the end vertex of this edge is the root of an infinite subtree each of whose vertices (including the root vertex) has at least one outgoing edge and one incoming edge of each type, see Fig.~\ref{f2}(b).
Note that the only formal difference between these definitions is the direction of the initial edge of type $i$. In the first tree the paths of type $i$ run from the root to infinity, while in the second tree they run in the opposite direction.

These definitions are standard for problems of this kind. In particular, they are similar to the definition of the corresponding probability for $k$-cores of undirected networks in terms of infinite ($k{-}1$)-ary trees, i.e., by definition, trees whose vertices have degrees at least $k$ (i.e. the minimal branching $k-1$) \cite{Dorogovtsev:dgm06,Goltsev:gdm06}.
The common instructive idea for these problems, is to introduce and define these probabilities and trees rooted at a randomly chosen edge (at one of its ends) in a way independent on the second end of the edge. Whether a tree, introduced in this definition, belongs to the giant component for such a problem or not, depends on the second end of the edge. In our problem, the first infinite tree (from the definition of $x_i$) surely belongs to the giant strongly viable component if on the second end of the edge there is an infinite tree of this kind (from the definition of $y_i$).  In this configuration clearly both these infinite trees completely belong to the giant strongly viable component. On the other hand, if the second end of the edge has no tree of this kind, then part of the first infinite tree (how large this fraction is depends on the structure of a network) can occur outside of the giant strongly viable component. The analytical construction, which we use, is not unique. The message passing technique, which is an equivalent approach to this class of problems, exploits the same ideas \cite{Bianconi,BianconiDorogovtsev}.

Importantly, even if the infinite tree is not entirely in the giant strongly viable component, it nonetheless leads to this component. Indeed, this specific tree (say, from the definition of $x_i$) is infinite, so the paths of type $i$ running from the root to infinity are infinite, and, in addition, for each vertex on such a path there is a finite probability to have at least one extra incoming edge with a $y_i$ tree on its second end. Consequently these infinite paths of type $i$ (one can say, this tree) lead to the giant strongly viable component.

The introduced probabilities play the role of the order parameters of the phase transition associated with the emergence of the giant strongly viable component and enable us to find its size and other structural characteristics for these multiplex networks.

Taking into account the locally tree-like structure of the networks, the probability that a randomly chosen edge of type $i$ comes in a vertex of degree $(\textbf{q}_{A}, \textbf{q}_{B},\ldots)$, or goes out of this vertex, is equal to $q_{in,i}P(\textbf{q}_A,\textbf{q}_B,\ldots)/\langle q_{in,i} \rangle$ and $q_{out,i}P(\textbf{q}_A,\textbf{q}_B,\ldots)/\langle q_{out,i} \rangle$, respectively.
The  self-consistency equations, represented schematically in Fig.~\ref{f2}, can be written for arbitrary $m$ as following:
\begin{eqnarray}
&&x_{i}=\sum_{\textbf{q}_A,\textbf{q}_B,\ldots}\frac{q_{in,i}P(\textbf{q}_A,\textbf{q}_B,\ldots)}{\langle
q_{in,i}\rangle}\Big[1-(1-x_i)^{q_{out,i}}\Big] \nonumber
\\[5pt]
&&\times\prod_{j\neq i}\Big[1-(1-x_j)^{q_{out,j}}\Big]\Big[1-(1-y_j)^{q_{in,j}}\Big]
, \label{eq1}
\\[5pt]
&&y_{i}=\sum_{\textbf{q}_A,\textbf{q}_B,\ldots}\frac{q_{out,i}P(\textbf{q}_A,\textbf{q}_B,\ldots)}{\langle
q_{out,i}\rangle} \Big[1-(1-y_i)^{q_{in,i}}\Big]\nonumber
\\[5pt]
&&\times\prod_{j\neq i}\Big[1-(1-x_j)^{q_{out,j}}\Big]\Big[1-(1-y_j)^{q_{in,j}}\Big]
.
\label{eq2}
\end{eqnarray}
Let us explain the right-hand terms in Eq.~(\ref{eq1}). For a vertex with $q_{out,i}$ outgoing edges of type $i$, probability that the second
end vertex of each of these outgoing edges is not the root of an infinite $x_i$ subtree 
is $(1-x_i)^{q_{out,i}}$. Consequently, $[1-(1-x_i)^{q_{out,i}}]$ is the probability that 
the second end vertex of at least one of these outgoing edges is the root of an infinite $x_i$ subtree.
Similarly, $\prod_{j\neq i}[1-(1-x_j)^{q_{out,j}}]$ is the probability that for each type $j\neq i$, at least one
of outgoing edges of this vertex of type $j$
leads to the $x_j$ subtrees.
Also, for a vertex with $q_{in,j}$ incoming edges of type $j\neq i$, $\prod_{j\neq i}[1-(1-y_j)^{q_{in}}]$, gives the probability that for each type $j$,
the second end vertex of at least one of these incoming edges is the root of an infinite $y_j$ subtree.
Similar arguments are valid for Eq.~(\ref{eq2}). The generating function technique \cite{NewmanStWatts} allows us to represent these equations in a more compact form (see Appendix). Equations~(\ref{eq1})--(\ref{eq2}) for probabilities $x_i$ and $y_i$ are written
for multiplex networks with edges of $m$ types, where $m\geq 1$. For the sake of simplicity, we consider $m=2$, and obtain the sizes of different components.


\section{Giant components in multiplex networks with two types of edges}

\label{two directed networks_sec}
\begin{figure*}
\begin{center}
\scalebox{0.38}{\includegraphics[angle=0]{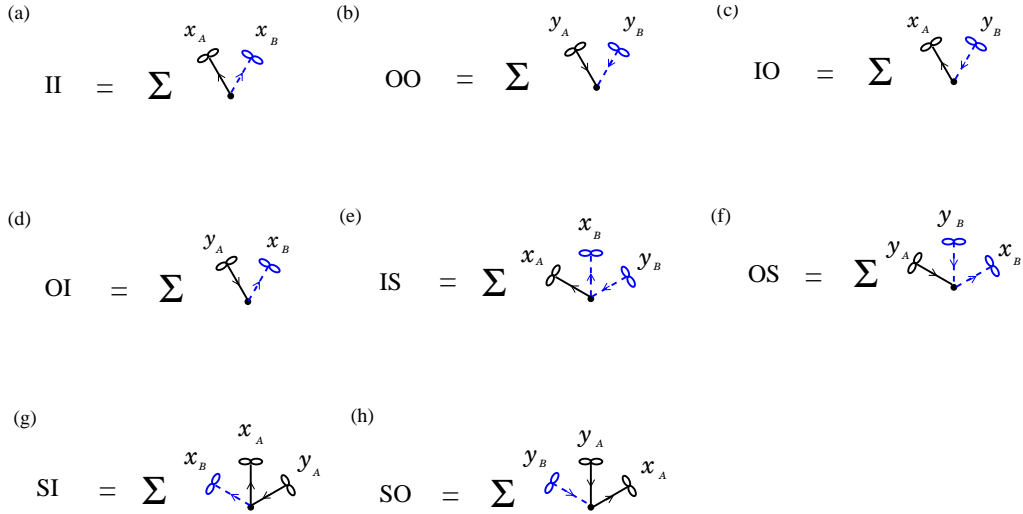}}
\end{center}
\caption{(Color online) Schematic representation of the probability that a vertex belongs to the following giant viable components: $(a)$ II, $(b)$ OO, $(c)$ IO, $(d)$ OI, $(e)$ IS, $(f)$ OS, $(g)$ SI, $(h)$ SO in the case of directed multiplex networks with edges of two colors, $A$ and $B$. For the analytical representation, see Eqs.~(\ref{eq9})--(\ref{eq16}).
}
\label{f4}
\end{figure*}

The probabilities $x_i$ and $y_i$ enable us to find the relative size of the giant strongly viable component exactly. Figure~\ref{f3} shows the probability that a vertex belongs to the giant strongly viable component, SS, in terms of the probabilities $x_i$ and $y_i$ for a multiplex network with two types of edges. Following the derivation of Eq.~(\ref{eq1}), we can obtain the relative size of this component:
\begin{widetext}
\begin{eqnarray}
SS=\sum_{\textbf{q}_A,\textbf{q}_B}P(\textbf{q}_A,\textbf{q}_B)\Big[1-(1-x_A)^{q_{out,A}}\Big]
\Big[1-(1-y_A)^{q_{in,A}}\Big]\Big[1-(1-x_B)^{q_{out,B}}\Big]\Big[1-(1-y_B)^{q_{in,B}}\Big]
.
\label{eq8}
\end{eqnarray}
\end{widetext}
Finding the sizes of the remaining giant viable components is a more difficult task than for SS, and we only estimate these sizes from below.
As we explained in Section~\ref{equations}, the infinite trees from the definitions of $x_i$ or $y_i$ lead to the giant strongly viable component. This enables us to indicate configurations based on these trees, which guarantee that a vertex belongs to a given giant component, see Fig.~\ref{f4}. Note that according to our definition, each of these giant components includes the giant strongly viable component. These combinations with the infinite trees $x_i$ and $y_i$, however, do not cover all possible configurations contributing to the probabilities that a vertex is in one of these giant components. The point is that the condition that each vertex in these trees has at least one outgoing edge and one incoming edge of each type is actually too strong, and some of their vertices can be allowed to have less connections of these types. This should results in larger sizes of these giant components than we estimate. Accounting for all these contributions turns out to be a difficult and challenging problem, demanding cumbersome calculations, so we have to leave it for a future work.

The analytical expression for II shown in Fig.~\ref{f4}(a) is as follows
\begin{eqnarray}
II\simeq\sum_{\textbf{q}_A,\textbf{q}_B}&&P(\textbf{q}_A,\textbf{q}_B)\Big[1-(1-x_A)^{q_{out,A}}\Big]
\nonumber
\\[5pt]
&&\times\Big[1-(1-x_B)^{q_{out,B}}\Big].
\label{eq9} \!\!\!\!\!
\end{eqnarray}
Similarly, for the out-out component, we use the probabilities $y_A$ and $y_B$, which
allows us to estimate the relative size of OO as
\begin{eqnarray}
OO\simeq\sum_{\textbf{q}_A,\textbf{q}_B}&&P(\textbf{q}_A,\textbf{q}_B)\Big[1-(1-y_A)^{q_{in,A}}\Big]
\nonumber
\\[5pt]
&&\times\Big[1-(1-y_B)^{q_{in,B}}\Big]
,
\label{eq10}
\end{eqnarray}
see Fig.~\ref{f4}(b).
Figs.~\ref{f4}(c) and (d) give the relative sizes of components IO and OI, respectively:
\begin{eqnarray}
IO\simeq\sum_{\textbf{q}_A,\textbf{q}_B}&&P(\textbf{q}_A,\textbf{q}_B)\Big[1-(1-x_A)^{q_{out,A}}\Big]
\nonumber
\\[5pt]
&&\times\Big[1-(1-y_B)^{q_{in,B}}\Big]
, \label{eq11}
\\
[5pt]
OI\simeq\sum_{\textbf{q}_A,\textbf{q}_B}&&P(\textbf{q}_A,\textbf{q}_B)\Big[1-(1-y_A)^{q_{in,A}}\Big]
\nonumber
\\[5pt]
&&\times\Big[1-(1-x_B)^{q_{out,B}}\Big]
. \label{eq12}
\end{eqnarray}
According to Figs.~\ref{f4}(e) and (f), the relative sizes of components IS and OS are, respectively,
\begin{eqnarray}
IS&\simeq&\sum_{\textbf{q}_A,\textbf{q}_B}P(\textbf{q}_A,\textbf{q}_B)\Big[1-(1-x_A)^{q_{out,A}}\Big]
\nonumber
\\[5pt]
&&\times\Big[1-(1-x_B)^{q_{out,B}}\Big]\Big[1-(1-y_B)^{q_{in,B}}\Big]
, \label{eq13}
\\
[5pt]
OS&\simeq&\sum_{\textbf{q}_A,\textbf{q}_B}P(\textbf{q}_A,\textbf{q}_B)\Big[1-(1-y_A)^{q_{in,A}}\Big]\nonumber
\\[5pt]
&&\times\Big[1-(1-x_B)^{q_{out,B}}\Big]\Big[1-(1-y_B)^{q_{in,B}}\Big]
. \label{eq14}
\end{eqnarray}
Finally, for relative sizes of components SI and SO, Figs.~\ref{f4}(g) and (h) give, respectively,
\begin{eqnarray}
SI&\simeq&\sum_{\textbf{q}_A,\textbf{q}_B}P(\textbf{q}_A,\textbf{q}_B)\Big[1-(1-x_B)^{q_{out,B}}\Big]
\nonumber
\\[5pt]
&&\times\Big[1-(1-x_A)^{q_{out,A}}\Big]\Big[1-(1-y_A)^{q_{in,A}}\Big]
, \label{eq15}
\\
[5pt]
SO&\simeq&\sum_{\textbf{q}_A,\textbf{q}_B}P(\textbf{q}_A,\textbf{q}_B)\Big[1-(1-y_B)^{q_{in,B}}\Big]\nonumber
\\[5pt]
&&\times\Big[1-(1-x_A)^{q_{out,A}}\Big]\Big[1-(1-y_A)^{q_{in,A}}\Big]
. \label{eq16}
\end{eqnarray}
Employing generating functions, one can represent Eqs.~(\ref{eq8})--(\ref{eq16}) in a more
compact form (see Appendix).

For the sake of simplicity we assume that there are no correlations between edges of different types and in- and out-degrees
for each type of edges $i$, i.e., $P(q_{in,i},q_{out,i})=P(q_{in,i})P(q_{out,i})$. Let us consider first the Erd\H{o}s--R\'enyi directed multiplex networks with Poisson in-degree and out-degree distributions,  $P(q_{in,i})=c_i^{q_{in,i}} e^{-c_i q_{in,i}}/q_{in,i}!$ and $P(q_{out,i})=c_i^{q_{out,i}} e^{-c_i q_{out,i}}/q_{out,i}!$, where $c_A$ and $c_B$ are the mean vertex degrees for edges $A$ and $B$, respectively. For the Poisson distribution, the generating function $G_i(x)$ and its first derivatives (see Appendix) are $G_i(x)= G_{1}^{in,i}(x)= G_{1}^{out,i}(x)= e^{-c_i (1-x)}$.

Inspecting Eqs.~(1)--(2) in the general case of $c_A\neq c_B$, we conclude that $x_i=y_i\equiv X_i$, and Eqs.~(\ref{eq1})--(\ref{eq2}) are simplified to the equation
\begin{equation}
X_i=(1-e^{-c_i X_i})(1-e^{-c_j X_j})^2
.
\label{eq17}
\end{equation}
The largest root of this equation plays the role of the order parameter in this problem. For the symmetric case $c_A=c_B\equiv c$, the non-zero solution exists only when $c$ exceeds the critical value $c=3.08912\ldots$.
\begin{figure}
\begin{center}
\scalebox{0.45}{\includegraphics[angle=0]{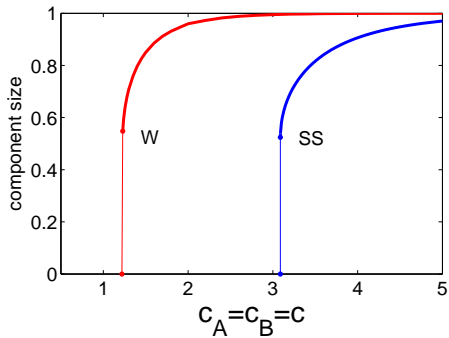}}
\end{center}
\caption{(Color online) The relative size of the giant strongly viable component (SS) of directed multiplex Erd\H{o}s--R\'enyi networks with two types of edges vs. mean in-, out-degree $c_A=c_B\equiv c$ of the networks. The W curve shows the dependence of the giant weakly viable component on $c$. This component emerges at $c=2.4554.../2$.
}
\label{f5}
\end{figure}

Furthermore, equations~(\ref{eq8})--(\ref{eq16}) provide the following expressions for the relative sizes of the giant viable components:
%
\begin{eqnarray}
&&
\!\!\!\!\!\!\!\!
SS=(1-e^{-c_A X_A})^2(1-e^{-c_B X_B})^2
,
\label{eq19}
\\[5pt]
&&\!\!\!\!\!\!\!\!
II=OO=IO=OI
\simeq(1{-}e^{-c_A X_A})(1{-}e^{-c_B X_B})
,
\label{eq20}
\\[5pt]
&&
\!\!\!\!\!\!\!\!
IS=OS\simeq X_A
,
\label{eq21}
\\[5pt]
&&
\!\!\!\!\!\!\!\!
SI=SO\simeq X_B
.
\label{eq22}
\end{eqnarray}
%
\begin{figure}
\begin{center}
\scalebox{0.44}{\includegraphics[angle=0]{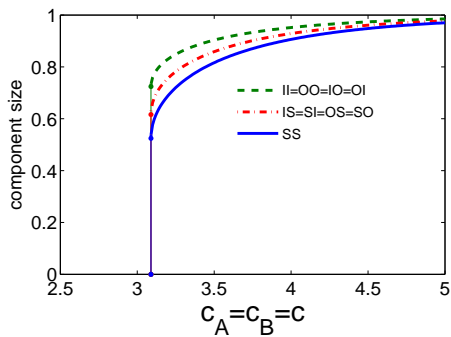}}
\end{center}
\caption{(Color online) The lower limit estimation of the sizes of different giant viable components compared with the size of the giant strongly component.
}
\label{f6}
\end{figure}

\begin{figure}
\begin{center}
\scalebox{0.44}{\includegraphics[angle=0]{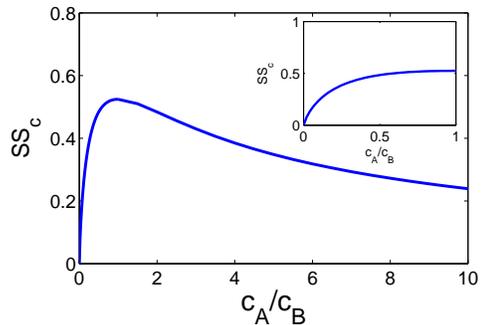}}
\end{center}
\caption{(Color online) The value of the discontinuity of
the giant strongly viable component vs. the ratio $c_A/c_B$ of the mean in- and out-degrees of networks $A$ and $B$.
}
\label{f7}
\end{figure}

\begin{figure}
\begin{center}
\scalebox{0.43}{\includegraphics[angle=0]{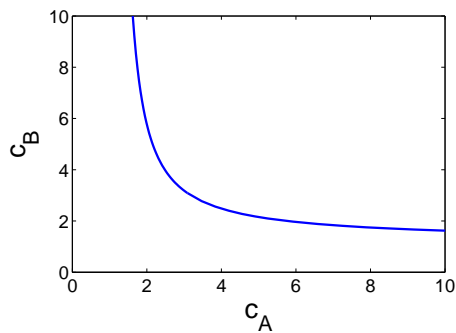}}
\end{center}
\caption{(Color online) Line on the plane $(c_A,c_B)$, which separates the phase with the giant strongly viable component from the normal phase.
}
\label{f8}
\end{figure}

For the case of $c_A=c_B=c$, in  Fig.~\ref{f5} the resulting dependence of the relative size of the giant strongly viable component on $c$ is compared with the corresponding dependence of the giant weakly viable component.
The giant weakly viable component is defined as the viable component of the undirected counterpart of the directed multiplex network, in which the directedness of the edges is ignored. This component emerges at $c=2.4554.../2$ \cite{SonChrisGrassPaczuski}. In Fig.~\ref{f6}, the lower limit estimates for the sizes of the other viable components compared with
SS are shown.
Figure~\ref{f7} demonstrates that the value of the jump at a critical point for SS is maximum when the mean degrees of the two networks coincide.
To obtain the critical point for the giant strongly viable component, we introduce $g_j(X_j)\equiv X_j-(1-e^{-c_j X_j})(1-e^{-c_i X_i})^2$. The critical point is determined by the condition $\det[\textbf{J}-\textbf{I}]=0$ for the Jacobian matrix $\textbf{J}$, defined as $J_{ij} = \partial f_j/\partial x_i$, and $\textbf{I}$ is the identity matrix. Substituting the resulting values $X_A$ and $X_B$ in Eq.~\ref{eq19}, we find the jump of the SS component at the transition point. The line of critical points for the SS component on the plane $(c_A,c_B)$ is shown in Fig.~\ref{f8}.

\begin{figure}
\begin{center}
\scalebox{0.43}{\includegraphics[angle=0]{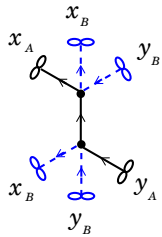}}
\end{center}
\caption{(Color online) Schematic representation of the probability that an edge of type $A$ belongs to the giant strongly viable component.
}
\label{f9}
\end{figure}

Our approach allows us to describe the structure of the viable components.
In particular, one can ask what is the probability that an edge belongs to the giant strongly viable component?
By definition, as is shown in Fig.~\ref{f9}, if an edge belongs to the strongly
viable component,
then
the end vertices of this edge should be connected to each other by paths of both colors running through infinity.
Hence the probability that a uniformly randomly chosen edge of type $i$ belongs to the giant strongly viable component is $P_i=x_i y_i$.

In many real systems there are more than two multiplex networks. In principle it is possible to generalize structure of viable components from two directed multiplex networks to a systems of $m$ directed multiplex networks. The main part, namely the strongly viable component of the network in which vertices are reachable from each other by directed paths of all $m$ types, is a subgraph of the strongly connected components of all networks.
Giant components in this work are classified according to the set of directed paths of all possible types from their vertices to the giant strongly viable component. For each type of edges, there are 3 possibilities: there is at least one path of this type coming from the giant strongly viable component to a vertex and no paths running through edges of this type in the opposite direction, there is at least one path of this type going to the giant strongly viable component from a vertex and no paths of this type running in the opposite direction, and, finally, there are paths of this type running in both directions. Hence one can find that the total number of viable components for a directed multiplex network with $m$ types of edges is $3^m$.


\section{Conclusions}

In this work we described the
topology of directed multiplex networks. We introduced a set of giant viable components for a directed multiplex network with two types of edges. The definitions of these components are based
on a set of directed paths between vertices which run following distinct types of edges. We showed how to find analytically the emergence points and the sizes of various giant viable components for directed multiplex networks with an  arbitrary joint in-, out degree distribution. We found that similar to a viable component in undirected multiplex networks, hybrid transitions occur at the points of emergence of the
giant viable connected components in directed multiplex networks.

In our analytical calculations, we considered uncorrelated and locally tree-like directed multiplex networks. We mostly focused on directed multiplex networks with two types of edges. In this case, the total number of giant viable components is $9$. In general, for the multiplex networks with $m$ types of edges, there are $3^{m}$ distinct viable components. 
For uncorrelated directed multiplex networks, we found exactly the size of the central, strongly viable component. We also estimated from below the sizes of the rest giant viable components. An exact calculation of the sizes and other characteristics of these components is a challenging task for a future work.

Our study has revealed an essentially more rich global organization of directed multiplex networks compared to single directed networks. We suggest that the knowledge
of the details of this complex structure will lead to a better understanding of processes taking place in these networks and of their function.

\begin{acknowledgments}
This work was partially supported by the projects FCT
PTDC/MAT/114515/2009 and PEst-C/CTM/LA0025/2011,
and the FET IP Project MULTIPLEX 317532.
\end{acknowledgments}


\appendix

\section{Generating function technique for directed multiplex networks}

The generating function $G(x,y)$ of
a given in-, out-degree distribution $P(q_{in},q_{out})$ for a directed network is
\begin{eqnarray}
G
(x,y)\equiv \sum_{q_{in},q_{out}} P(q_{in},q_{out})x^{q_{out}}y^{q_{in}}
.
\label{eq3}
\end{eqnarray}
Using the function $G(x,y)$, one can obtain the generating functions $G_{1}^{in}(x)$ and $G_{1}^{out}(y)$ of the joint in-, out-degree distribution of the end vertex of a randomly chosen incoming or outgoing edge, respectively:
\begin{eqnarray}
G_{1}^{in}(x)&=&\frac{\partial_{y}G(x,y)|_{y=1}}{\partial_{y}G(x,1)|_{y=1}}
,
\label{eq4}
\\[5pt]
G_{1}^{out}(y)&=&\frac{\partial_{x}G(x,y)|_{x=1}}{\partial_{x}G(1,y)|_{x=1}}
.
\label{eq5}
\end{eqnarray}
If we assume that there is no correlation between the degrees $q_A, q_B, \ldots$ of a vertex, so that $P(\textbf{q}_A,\textbf{q}_B) = P(\textbf{q}_A)P(\textbf{q}_B)\ldots$, then using these definitions, Eqs.~(\ref{eq1})--(\ref{eq2}) can be
represented as
\begin{eqnarray}
x_{i}&=&\Big[1-G_{1}^{in,i}(1-x_{i})\Big]\prod_{j\neq i}\Big[1-G^{j}(1-x_{j},1)
\nonumber
\\[5pt]
&-&G^{j}(1,1-y_{j})+G^{j}(1-x_{j},1-y_{j})\Big],
\label{eq6}
\\[5pt]
y_{i}&=&\Big[1-G_{1}^{out,i}(1-y_{i})\Big]\prod_{j\neq i}\Big[1-G^{j}(1-x_{j},1)
\nonumber
\\[5pt]
&-&G^{j}(1,1-y_{j})+G^{j}(1-x_{j},1-y_{j})\Big],
\label{eq7}
\end{eqnarray}
where the index $i$
refers to
types of edges $A,B,\ldots$.

The relative sizes of the giant viable components can be written in terms of generating functions as follows:
\begin{eqnarray}
&&II=\Big[1-G^{A}(1-x_{A},1)\Big]\Big[1-G^{B}(1-x_{B},1)\Big],
\label{eq4aA}
\\[5pt]
\!\!\!\!\!
&&OO=\Big[1-G^{A}(1,1-y_{A})\Big]\Big[1-G^{B}(1,1-y_{B})\Big] ,
\\[5pt]
\!\!\!\!\!
&&IO=\Big[1-G^{A}(1-x_{A},1)\Big]\Big[1-G^{B}(1,1-y_{B})\Big] ,
\\[5pt]
\!\!\!\!\!
&&OI=\Big[1-G^{A}(1,1-y_{A})\Big]\Big[1-G^{B}(1-x_{B},1)\Big] ,
\\[5pt]\nonumber
\!\!\!\!\!
\end{eqnarray}
\begin{widetext}
\begin{eqnarray}
&&IS=\Big[1-G^{A}(1-x_{A},1)\Big]\Big[1-G^{B}(1-x_{B},1)-G^{B}(1,1-y_{B})+G^{B}(1-x_B,1-y_{B})\Big]
,
\\[5pt]
&&SI=\Big[1-G^{A}(1-x_{A},1)-G^{A}(1,1-y_{A})+G^{A}(1-x_A,1-y_{A})\Big]\Big[1-G^{B}(1-x_{B},1)\Big]
,
\\[5pt]
&&OS=\Big[1-G^{A}(1,1-y_{A})\Big]\Big[1-G^{B}(1-x_{B},1)-G^{B}(1,1-y_{B})+G^{B}(1-x_B,1-y_{B})\Big]
,
\\[5pt]
&&SO=\Big[1-G^{A}(1-x_{A},1)-G^{A}(1,1-y_{A})+G^{A}(1-x_A,1-y_{A})\Big]\Big[1-G^{B}(1,1-y_{B})\Big]
,
\\[5pt]
&&\!\!\!\!\!\!\!\!\!\!\!\!SS{=}\Big[\!1{-}G^{A}(1{-}x_{A},1){-}G^{A}(1,1{-}y_{A}){+}G^{A}(1{-}x_A,1{-}y_{A})]\Big]\!
\Big[1{-}G^{B}(1{-}x_{B},1){-}G^{B}(1,1{-}y_{B}){+}G^{B}(1{-}x_B,1{-}y_{B})]\Big]
\!. \label{eq4kA}
\end{eqnarray}
%
\end{widetext}



\end{document}